\documentstyle [12pt]{article}
\newcommand{\be}{\begin{equation}}
\newcommand{\ee}{\end{equation}}
\newcommand{\ra}{\rightarrow}

\newcommand{\lsim}{\stackrel{<}{\sim}}
\newcommand{\qnw}{ Q^{(NW)} }
\newcommand{\qk}{ Q^{(\kappa)} }
\newcommand{\pll}{L_{\rm Pl}}
\newcommand{\plm}{M_{\rm Pl}}
\begin{document}
\begin{titlepage}
\begin{flushright}
 IFUP-TH 19/93\\
 May 1993\\ (revised Nov. 1993)\\
 hep-th/9305163
\end{flushright}

\vspace{8mm}

\begin{center}

{\Large\bf  Quantum Groups, Gravity and the

\vspace{4mm}

Generalized Uncertainty Principle }\\

\vspace{12mm}
{\large Michele Maggiore}

\vspace{3mm}

I.N.F.N. and Dipartimento di Fisica dell'Universit\`{a},\\
piazza Torricelli 2, I-56100 Pisa, Italy.\\
\end{center}

\vspace{4mm}

\begin{quote}
ABSTRACT. We investigate the relationship between the
 generalized uncertainty principle in quantum gravity
and   the quantum deformation of the Poincar\'e
algebra.
We find
that a deformed Newton-Wigner position operator and the
generators
of spatial translations and rotations of the deformed
Poincar\'e algebra
obey a deformed Heisenberg algebra from which the
generalized
uncertainty principle follows. The result indicates that
in the $\kappa$-deformed
Poincar\'e algebra a minimal observable length emerges
naturally.
\end{quote}
\end{titlepage}
\clearpage

\section{Introduction}

There are many indications that in quantum gravity there
might exist a minimal
observable
distance on the order of the Planck length. The emergence
of a minimal
length is usually  considered a {\em dynamical} phenomenon,
related to the fact that at the Planck scale  there are
violent
fluctuations of the metric and even topology changes, as in
Wheeler spacetime
foam~\cite{Whe,GH}. In the context of string theories, the
emergence
of a minimal measurable distance is nicely encoded in
 a generalized uncertainty principle [3-9]:
\be\label{1}
\Delta x\ge \frac{\hbar}{\Delta p} +\alpha G\Delta p\,
\ee
where $\alpha$ is a constant. (We have written explicitly
$\hbar$ and $G$ and we have set $c=1$).
Eq.~(1) has been  obtained from the study of
string collisions
at planckian energies, so again it has a {\em dynamical}
origin
(although for strings the dynamical and kinematical
aspects are strongly correlated).
The purpose
of this paper is to  investigate whether  it is possible
to understand eq.~(1) at a purely
{\em kinematical} level, independently of any specific dynamical
theory.

Our original motivation for this investigation is the fact
that in ref.~\cite{MM} we   obtained eq.~(1) without considering
strings, but rather discussing a Gedanken experiment
in which the
radius of the apparent horizon of a black hole is measured.
In this context the generalized
uncertainty principle is rediscovered
using only very general and model-independent
considerations which would
presumably be fulfilled by
any candidate quantum theory of gravitation. As
a matter of fact, the only  physical input is
the existence of Hawking
radiation~\cite{Haw} emitted by black holes.
This fact suggests to look
for a mathematical structure which reproduces
eq.~(1) in a natural way. In ref.~\cite{MM3}
we have indeed found that a suitable algebraic
structure exists,
and it is given by the deformed Heisenberg algebra
\begin{eqnarray}
\left[ X_i ,X_j \right] &= & -\frac{\hbar^2}{4\kappa^2}
i\epsilon_{ijk}J_k\label{xx}\\
\left[ X_i , P_j \right]   &= & i\hbar\delta_{ij}\,
\left( 1+\frac{ {\bf  P}^2+m^2}{4\kappa^2}\right)^{1/2}\, .
\label{xp}
\end{eqnarray}
The commutation relations of the angular momentum $J_i$ with the
coordinates $X_i$ and the momenta $P_i$, as well as between
themselves, are the standard ones, $\left[ X_i,J_j\right] =i
\epsilon_{ijk}X_k$, etc. The deformation parameter $\kappa$ has
dimensions of mass, and in the
limit $\kappa\ra\infty$ the undeformed
Heisenberg algebra is recovered. In the following
we will identify $\kappa$ with the Planck mass, times a numerical
constant.\footnote{A comment on the conventions is in order.
In the above formulas
the deformation parameter always appears in the combination
$2\kappa$. Because of this, in~\cite{MM3} we have rescaled $\kappa$
by a factor of 2.
In this paper we do not perform such a rescaling,
so that the parameter that we call $\kappa$
here agrees with the one
used in the
literature on the $\kappa$-deformed Poincar\'e algebra.}
The algebra defined by eqs.~(\ref{xx},\ref{xp})
is well defined, in the
sense that the Jacobi identities are satisfied. Moreover, we
 found in~\cite{MM3} that the requirement that Jacobi
identities are satisfied is so restrictive that,
within rather reasonable assumptions,
this is the unique possible
deformation of the Heisenberg algebra, when the deformation
parameter is dimensionful.

{}From eq.~(\ref{xp}) the generalized uncertainty principle
follows:
\be\label{gup}
\Delta x_i\Delta p_j\ge\frac{\hbar}{2}\delta_{ij}\langle
\left( 1+\frac{ {\bf  P}^2+m^2}{4\kappa^2}\right)^{1/2}
\rangle\, .
\ee
(Here and in the following we denote operators
by capital letters and their expectation values with
small case letters). Expanding the square root  at lowest
order and using $\langle {\bf P}^2\rangle ={\bf p}^2
+(\Delta p)^2$ we find
\be
\Delta x_i\Delta p_j\ge\frac{\hbar}{2}\delta_{ij}
\left( 1+\frac{{\bf p}^2+m^2+(\Delta p)^2}{8\kappa^2}
\right)\, .
\ee
Thus, in
the regime ${\bf p}^2+m^2\ll\kappa^2, \Delta p\lsim\kappa$ we
recover  eq.~(\ref{1}). Instead, in the asymptotic regime
${\bf p}^2\sim (\Delta p)^2\gg\kappa^2$ eq.~(\ref{gup}) gives
\be
\Delta x\ge {\rm const.}\times\frac{\hbar}{\kappa}\, .
\ee
The purpose of this paper is to
illustrate the relationship between
the generalized uncertainty principle and
the quantum deformation of
the Poincar\'e
algebra. This investigation can be useful because, on the
one hand, we can find a kinematic framework in which
eqs.~(\ref{xx},\ref{xp}) are
 satisfied. Independently of
whether this specific framework will be  relevant or not for
quantum gravity, we can expect to gain a better understanding of
the physical meaning of
the deformed Heisenberg algebra and of the generalized
uncertainty principle.
 On the other
hand, such an investigation can be interesting from the
point of view
of quantum groups [13-18],
since it indicates that in such a structure a minimum length
is automatically build in (at least when the deformation
parameter is dimensionful).

The problem of
finding a quantum deformation of the Poincar\'e group has
received much attention
recently, and different approaches have been
developed. An important line of research is concerned with
defining the differential calculus on
quantum groups [18-21]; then
one can define
curvatures through Cartan's equation, and try to construct
a $q$-generalization of Einstein action.

A second approach
consists in looking for a deformation of the algebra,
rather than of the group [22-26].
A very interesting technique which has been used in this context is
the contraction procedure first introduced in~\cite{Fi}. One
first considers the $q$-deformation of  the anti-de~Sitter algebra,
$U(O(3,2))$. This can be done with the standard Drinfeld-Jimbo
method~\cite{Dri,Jim},
and introduces a dimensionless deformation parameter $q$.
Then one sends to infinity the  de~Sitter radius $R$ while
$q\ra 1$ in  such a way that $R\log q\ra\kappa^{-1}$, fixed.  One
therefore recovers a
deformation of the $d=4$ Poincar\'e algebra which depends
on a {\em dimensionful} parameter $\kappa$. In this way
 a fundamental length
enters the theory.

For our purposes, what is needed is the knowledge of the
deformed algebra.
We will consider
the deformed Poincar\'e algebra given in~\cite{Luk3};
however, our line of
reasoning is more general,
and could be adapted to different deformations,
as long as they introduce a {\em dimensionful} parameter.

In the following, a fundamental role is played by the
Newton-Wigner position
operator~\cite{NW}. In the undeformed case it
represents the relativistic position operator
 of a particle.  The
main concern of this
paper will be to find a proper generalization of this
operator to the deformed case.
The plan of the paper is as follows.
In sect.~2 we recall the main results
concerning the quantum Poincar\'e
algebra which will be useful in the
following. In sect.~3 we discuss the
generalization of  the Newton-Wigner
position operator to the deformed case and
in sect.~4 we discuss  our results.

\section{The quantum Poincar\'e algebra}

We now briefly recall the main properties of
 the $\kappa$-deformed Poincar\'e algebra given
in~\cite{Luk3} (see also~\cite{Luk1,Luk2,Gil}).
All commutators are the same as in the usual Poincar\'e algebra,
except for the boost--boost and boost--3-momentum commutators:
\be\label{KK}
\left[ K_i,K_j\right] =
-i\epsilon_{ijk}\left( J_k\cosh\frac{P_0}{\kappa}
-\frac{1}{4\kappa^2}P_k {\bf P\cdot J}\right)\, ,
\ee
\be\label{KP}
\left[ P_i,K_j\right] =-i\delta_{ij}\kappa\sinh
\frac{P_0}{\kappa}\, .
\ee
Here $P_{\mu},J_i,K_i$ are the deformed four-momentum,
angular momentum and boost generators, respectively.
In the limit $\kappa\ra\infty$ the standard commutators
are recovered.
The first Casimir operator is~\cite{Gil,Luk3}
\be
C_1={\bf P}^2-\left( 2\kappa\sinh\frac{P_0}{2\kappa}\right)^2\, ,
\ee
so that the dispersion relation reads
\be\label{disp}
\left( 2\kappa\sinh\frac{P_0}{2\kappa}\right)^2=m^2+{\bf P}^2\, .
\ee
To study unitary representations, one considers the Hilbert
space with
a positive definite scalar product invariant under
$\kappa$-deformed Poincar\'e transformations,
\begin{eqnarray}
(\phi,\psi ) & = & \int
\frac{d^4p}{(2\pi )^4}\,\theta (p_0)2\pi
\delta \left( {\bf p}^2 +m^2
- 4\kappa^2\sinh^2\frac{p_0}{2\kappa}\right)
\phi^{\ast}(p)\psi (p)\, \nonumber\\ \label{ps2}
  & = &
\int\,\frac{d^3p}{(2\pi )^3 2 \kappa \sinh (p_0/\kappa ) }
\phi^{\ast}({\bf p})\psi ({\bf p})\, ,
\end{eqnarray}
where in
the last line $p_0$ has become a notation for the positive
solution of eq.~(\ref{disp}). This scalar product
has the correct limit for $\kappa\ra\infty$.
Limiting
ourselves  to the spin zero
case, so  that the term
${\bf P\cdot J}$ in eq.~(\ref{KK}) does not
contribute,\footnote{In ref.~\cite{MM3} a strong
restriction on the possible forms of the deformed algebra was
obtained requiring that the Jacobi identities are satisfied
independently of whether ${\bf p\cdot J}=0$ or not.
Limiting ourselves to the case
${\bf p\cdot J}=0$ we might in principle introduce some extra
solution. We will see
however that our final result is the same as the
one found in~\cite{MM3}.}
the representation of the
generators of the
deformed Poincar\'e algebra on this Hilbert space reads
\begin{eqnarray}
P_{\mu} &=& p_{\mu}\nonumber\\
J_i     &=&-i\epsilon_{ijk}p_j\frac{\partial}{\partial p_k}\\
K_i     &=&i\kappa\sinh
(\frac{p_0}{\kappa}) \frac{\partial}{\partial p_i}
\nonumber\, .
\end{eqnarray}
These operators are hermitean with respect to the scalar
product~(\ref{ps2}).

\section{The Newton-Wigner position operator}

\subsection{The undeformed case}

We now wish to
represent the relativistic position operator
on our Hilbert
space. Let us first recall  how this is done
in the undeformed case.
The  concept of relativistic position operator
was first introduced in a fundamental paper  by Newton and
Wigner~\cite{NW},
and further discussed by Wightman~\cite{Wig} and
Mackey~\cite{Mac}.
(For a pedagogical discussion see also~\cite{BR}).
For a massive
particle, one  considers the Hilbert space of functions with
the invariant scalar product
\be\label{ps}
(\phi,\psi )=\int\frac{d^3p}{(2\pi )^3 2p_0}
\phi^{\ast}({\bf p})\psi ({\bf p})\, ,
\ee
where $p_0$ is the positive solution of $p^2=m^2$.
On this
space, the
generators of the Poincar\'e group have the well-known
realization (limiting ourselves again to the spin zero case)
\begin{eqnarray}
P_{\mu} &=& p_{\mu}\nonumber\\
J_i     &=&-i\epsilon_{ijk}p_j\frac{\partial}{\partial p_k}\\
K_i     &=&ip_0\frac{\partial}{\partial p_i}\nonumber\,\, .
\end{eqnarray}
The representation of the
Newton-Wigner position operator $Q_i^{(NW)}$ is
\be\label{15}
\qnw_i=i\hbar
\left(\frac{\partial}{\partial p_i}-\frac{p_i}{2p_0^2}\right)\, .
\ee
It satisfies the commutation relations
\begin{eqnarray}
\left[ \qnw_i ,\qnw_j \right] &= & 0 \\
\left[ \qnw_i , P_j \right]   &= & i\hbar\delta_{ij}\\
\left[ \qnw_i,J_j \right] &=&i\epsilon_{ijk}\qnw_k\, .
\end{eqnarray}
The second term on the right-hand side of eq.~(\ref{15}) is
chosen in such a way that $\qnw_i$
 is hermitean with the scalar product given in eq.~(\ref{ps}).
In the non-relativistic
limit ${\bf Q}^{(NW)} =\hbar{\bf K}/m$. The time
derivative of $\qnw$ in the Heisenberg representation is
\be
\frac{d}{dt}\qnw_i
=\frac{i}{\hbar}\left[ P_0,\qnw_i\right] =
\frac{p_i}{p_0}\, ,
\ee
where the
relation $\partial p_0/\partial p_i=p_i/p_0$ has been used, since
we are working on-mass-shell. Therefore, the time derivative of
$\qnw_i$ is actually the relativistic
velocity of the particle, which is a necessary
requirement if we want to identify it with the position operator.

\subsection{The deformed case}

We must now find an operator  $\qk_i$ which  generalizes the
Newton-Wigner position operator to the $\kappa$-deformed case.
Two necessary requirements are,
first, that it should
reduce to $\qnw_i$ in the $\kappa\ra\infty$ limit and,
second, that it should be hermitean with respect
to the scalar product given in eq.~(\ref{ps2}).
It is also natural to ask
that the commutation relations with ${\bf J}$ are
not modified, so that it remains a vector under
space rotations. Then the
operator must be of the general form
\be
\qk_i=i\hbar
\left(A(\frac{p_0}{\kappa})\frac{\partial}{\partial p_i}-
             B(\frac{p_0}{\kappa})\frac{p_i}{2p_0^2}\right)\, ,
\ee
with $A(0)=B(0)=1$. The hermiticity condition gives immediately
\be\label{B}
B(\frac{p_0}{\kappa})=\frac{p_0^2}{\kappa\sinh p_0/\kappa}
\left[ \frac{1}{\kappa}\coth(\frac{p_0}{\kappa})
A(\frac{p_0}{\kappa})
-\frac{dA}{dp_0}\right]\, .
\ee
In terms of
the function $A$ one computes the following commutators,
\begin{eqnarray}
\left[ \qk_i ,\qk_j \right]
&= & -\frac{\hbar^2A}{\kappa\sinh (p_0/\kappa)}
\frac{dA}{dp_0} i\epsilon_{ijk}J_k\\
\left[ \qk_i , P_j \right]   &= & i\hbar A\delta_{ij}\, .\\
\end{eqnarray}
It would be
tempting at this stage to set $A(\frac{p_0}{\kappa})=1$;
the corresponding operator
\be\label{qk}
\qk_i=i\hbar\left(\frac{\partial}{\partial p_i}-
\frac{\cosh (p_0/\kappa )}
{2\kappa^2\sinh^2(p_0/\kappa )}p_i\right)\, ,
\ee
is hermitean with
respect to the scalar product ~(\ref{ps2}) and satisfies
\be\label{undef}
\left[ \qk_i ,\qk_j \right] =0,\hspace{5mm}
\left[ \qk_i ,P_j \right] = i\delta_{ij}\, .
\ee
Before interpreting
it as the generalization of the position operator
to the $\kappa$-deformed case, we must however
check if its time derivative
is the velocity of the particle.

A priori we do
not know how to define the velocity in terms of energy and
momentum in the deformed case. In principle, the relation
${\bf p}=p_0{\bf v}$
can be modified. If we take the time derivative
of the operator $\qk$ given in eq.~(\ref{qk}) we find
\be\label{v1}
\dot{Q}^{(\kappa )}_i = \frac{i}{\hbar}
 \left[ P_0,\qk_i\right] =
 \frac{p_i}{ \kappa\sinh (p_0/\kappa ) }
=\frac{p_i}{\sqrt{m^2+{\bf p}^2} \cosh (p_0/2\kappa )}\, ,
\ee
where we have used the dispersion relation~(\ref{disp}).
We see that,
for a particle with mass $m$, $|\dot{Q}^{(\kappa )}|$
is bounded  by
\be
|\dot{Q}^{(\kappa )}_i|\leq \frac{1}{\cosh (p_0/(2\kappa) )}
<\left( 1+\frac{m^2}{4\kappa^2}\right)^{-1/2}\, ,
\ee
instead of beeing
allowed to vary between zero and one as we expect for a
velocity. Furthermore, as a function of $p_0$,
it  reaches a maximum value
smaller than one, and then decreases exponentially
for large $p_0$. Even if
our understanding of
physics at the Planck scale is limited, such a
behavior seems rather non-sensical,
and suggests that one cannot
 identify the
right-hand side of
eq.~(\ref{v1}) with the velocity of a particle. In turns,
this means that the operator $\qk_i$, which satisfies the
undeformed commutation relations~(\ref{undef}),
cannot represent the relativistic
position operator in the $\kappa$-deformed theory.

We therefore need a
criterium which allows us to identify the velocity
operator. We suggest that the proper deformation
of the  relation between
momentum, velocity and energy is
\be\label{vel}
p_i=2\kappa\sinh (\frac{p_0}{2\kappa})v_i\, .
\ee
This assumption is rather natural, since it just amounts to the
replacement
$p_0\ra 2\kappa\sinh (\frac{p_0}{2\kappa})$, which is the  same
that takes place in the Casimir operator.
The relation (\ref{vel}) has the correct undeformed limit, and
$v$ is allowed
to vary between zero and one and is a monotonic function of
energy. In fact, eliminating
$\sinh (\frac{p_0}{2\kappa})$
with the use of the dispersion relation, one
finds
\be\label{pv}
p_i=\gamma mv_i, \hspace{15mm}\gamma = (1-v^2)^{-1/2}\, ,
\ee
so that this classical relation is not deformed.

It is easy to see that, if we require the time
derivative of the
position operator to be $p_i/(2\kappa\sinh\frac{ p_0}{2\kappa} )$,
we get
\be
A(\frac{p_0}{\kappa})=\cosh\frac{p_0}{2\kappa}\, .
\ee
The function $B$ then follows from eq.~(\ref{B}).
We are
therefore lead to propose the following generalization of the
Newton-Wigner position operator, which we denote $X_i$:
\be
X_i=i\hbar\cosh\frac{p_0}{2\kappa}
\left(\frac{\partial}{\partial p_i}-
\frac{p_i}{8\kappa^2\sinh^2(p_0/2\kappa)}\right)\, ,
\ee
which
is hermitean with respect to the scalar product~(\ref{ps2}),
 has the classical commutation relations with $J_i$
and satisfies $\dot{X}=v$, with $v$ defined by eq.~(\ref{pv}).
It is now
straightforward to compute the    $\left[ X,X\right]$ and
$\left[ X,P\right]$ commutators:
\begin{eqnarray}
\left[ X_i ,X_j \right] &=
& -\frac{\hbar^2}{4\kappa^2}i\epsilon_{ijk}J_k\\
\left[ X_i , P_j \right]   &= & i\hbar\delta_{ij}
\cosh\frac{P_0}{2\kappa}\, .
\label{comm}
\end{eqnarray}
Using the
dispersion relation, eq.~(\ref{disp}), we see that this is
just the algebra given in eqs.~(\ref{xx},\ref{xp}). Note that
eq.~(\ref{xp}) is written in a form which is
independent of
the specific dispersion relation. In the $\kappa$-Poincar\'e
algebra
it takes the form~(\ref{comm}), but we can as well consider
eqs.~(\ref{xx},\ref{xp}) within the standard  Poincar\'e group,
and then ${\bf P}^2+m^2=E^2$.

The result that
we have obtained does not come out as a surprise,
since we have shown in ref.~\cite{MM3} that the
$\kappa$-deformation of the
Heisenberg algebra is (essentially\footnote{In
ref.~\cite{MM3} we also found
a second  possible solution of the Jacobi
identities, of the form $\left[ X_i,P_j\right] =i\hbar
\delta_{ij}\{ 1-({\bf P}^2+m^2)/(4\kappa^2)\}^{1/2}$.
It is easy to see that one obtains
 this algebra using the $\kappa$-deformed Poincar\'e algebra
suggested in refs.~\cite{Luk1,Gil} instead of the one suggested
in ref.~\cite{Luk3}, since the former can be obtained from
the latter with the formal replacement $\kappa\ra i\kappa$.})
unique.

The fact that $\left[ X_i,X_j\right]$ is non zero
is consistent with the spirit
of non-commutative
geometry~\cite{Connes}, which is at the basis of the
quantum group approach to physics at the Planck scale~\cite{Maj}.
The non-commutativity shows up only at length scales on
the order of
the Planck length. It is also important to observe that
the deformed
Heisenberg algebra ties the generalized uncertainty principle with
non-commutativity of space-time at very short distances.

The deformation constant $\kappa$ can be estimated
if we assume that the uncertainty principle obtained  from
quantum groups at lowest order in $\Delta p/\kappa$
and $E\ll\kappa$,
which for $i=j$ reads
\be
\Delta x\Delta p\ge\frac{\hbar}{2} (1+\frac{(\Delta p)^2}{8\kappa^2}
+\ldots )\, ,
\ee
agrees with the one found
in string theory, which  reads
(apart from numerical constants of order one)
\be
\Delta x\Delta p\ge\frac{1}{2} (\hbar +\alpha '(\Delta p)^2)\, .
\ee
Here $\alpha '$ is the inverse
string tension, $\alpha '=\lambda_s^2/(2\hbar )$,
and $\lambda_s$
is the quantization constant of string theory; its relation
to the Planck length $\pll$ is somewhat model
dependent. In heterotic string
theory $\pll =\alpha_{\rm GUT}\lambda_s/4$.
In this case, therefore,
the comparison suggests
\be\label{het}
\kappa\sim\frac{1}{8}\alpha_{\rm GUT}\plm\sim
\left( 10^{-2}-10^{-3}\right)\plm\, .
\ee

\section{Discussion}

We have found
that the $\kappa$-deformed Poincar\'e algebra provides
an explicit realization of the $\kappa$-deformed Heisenberg
algebra, eqs.~(\ref{xx},\ref{xp}), once $X$ is
identified with a suitably
deformed
Newton-Wigner position operator. Other  definitions of
the deformed Newton-Wigner position operator are possible
(a different definition is
suggested in~\cite{Bac}). This ambiguity is  due to the fact that
the definition of velocity in terms of energy and momentum in the
$\kappa$-deformed theory is not fixed a priori, the only
requirement beeing that it should reduce to the classical relation
as $\kappa\ra\infty$. Our definition is dictated by the choice
that the relation $p=\gamma mv$ is not deformed, see eq.~(\ref{pv}).

A very relevant feature of
the $\kappa$-deformed Heisenberg algebra is the fact that
it is not compatible with exact  Lorentz invariance at the
Planck scale, as it is clear from the fact that it implies
the existence of a minimal {\em spatial\/ }
 length --~a concept which is
obviously non Lorentz-invariant. The fact that at the Planck scale
Lorentz boost should saturate has been suggested recently
by Susskind~\cite{Sus}.
The $\kappa$-Poincar\'e group provides
an explicit example of a kinematical framework in which
Lorentz transformations are modified. In this case
Lorentz invariance is  broken by the parameter $\kappa$.
Note also that there is no $\kappa$-deformed Lorentz subalgebra
of the $\kappa$-Poincar\'e algebra, since the boost-boost
commutator  involves the momentum.
 This explicit example also shows clearly
how a fundamental length can emerge at a purely kinematical level.

The  fact that quantum groups
can  provide the kinematical framework of physical
system was already realized in~\cite{Fi2,Fi3}.
The authors of this very interesting work
consider the propagation of phonons in a harmonic crystal
in 1+1 dimensions, and discover that the $d=2$
$\kappa$-deformed Poincar\'e
algebra is its kinematical
symmetry. Our approach could be considered
complementary to theirs, since  we rather
start with a $\kappa$-deformed algebra and
discover that a minimal length emerges.

The emergence
of a minimal length obviously has  important consequences
also
concerning the possibility that quantum groups provide a natural
ultraviolet cutoff mechanism for quantum field theory~\cite{Maj2}.

Finally, we note that  eq.~(\ref{gup}) agrees with eq.~(1) only
at lowest order in $\Delta p/\kappa$. In particular,
asymptotically
eq.~(\ref{gup})
gives $\Delta x\ge \hbar/\kappa$, while
eq.~(1) gives
$\Delta x\ge \hbar\Delta p/\kappa^2$.
It is easy to see why the
arguments presented in~\cite{MM}
fail in the region $\Delta p\gg\kappa$.
In our Gedanken experiment $\Delta p$
was on the order of the energy of the
particle used to probe the
black hole; and we cannot treat
semiclassically a particle with
super-planckian energy. It would  be
interesting to see if higher order
terms in $\Delta p/\kappa$ can be
obtained in the string theoretic derivation of eq.~(1).

\vspace{5mm}

Acknowledgments. I thank L.~Castellani, K.~Konishi, M.~Shifman,
A.~Vainshtein and G.~Vitiello for useful
discussions. I especially thank  M.~Mintchev
for fruitful discussions.

\end{document}